# Generation of photoionized plasmas in the laboratory of relevance to accretion-powered x-ray sources using keV line radiation

D. Riley[1,2], R.L. Singh[2,3], S White[2], M. Charlwood[2], D. Bailie[2], C. Hyland[2], T. Audet[2], G. Sarri[2], B. Kettle[4], G. Gribakin[2], S.J. Rose[4], E.G. Hill[4], G.J. Ferland[5], R.J.R. Williams[6] F.P. Keenan[2]

## Abstract

We describe laboratory experiments to generate x-ray photoionized plasmas of relevance to accretion-powered x-ray sources such as neutron star binaries and quasars, with significant improvements over previous work. A key quantity is referenced, namely the photoionization parameter, defined as $\xi = 4\pi F/n_e$ where $F$ is the x-ray flux and $n_e$ the electron density. This is normally meaningful in an astrophysical steady-state context, but is also commonly used in the literature as a figure of merit for laboratory experiments that are, of necessity, time-dependent. We demonstrate emission-weighted values of $\xi > 50$ erg-cm s$^{-1}$ using laser-plasma x-ray sources, with higher results at the centre of the plasma which are in the regime of interest for several astrophysical scenarios. Comparisons of laboratory experiments with astrophysical codes are always limited, principally by the many orders of magnitude differences in time and spatial scales, but also other plasma parameters. However useful checks on performance can often be made for a limited range of parameters. For example, we show that our use of a keV line source, rather than the quasi-blackbody radiation fields normally employed in such experiments, has allowed the generation of the ratio of inner-shell to outer-shell photoionization expected from a blackbody source with ~keV spectral temperature. We compare calculations from our in-house plasma modelling code with those from Cloudy and find moderately good agreement for the time evolution of both electron temperature and average ionisation. However, a comparison of code predictions for a K-β argon X-ray spectrum with experimental data reveals that our Cloudy simulation overestimates the intensities of more highly ionised argon species. This is not totally surprising as the Cloudy model was generated for a single set of plasma conditions, while the experimental data are spatially integrated.

[1] Corresponding author d.riley@qub.ac.uk
[2] School of Mathematics and Physics, Queen's University Belfast, University Road, Belfast BT7 1NN, UK
[3] ELI Beamlines Centre, Institute of Physics of the Czech Academy of Sciences, 252 41 Dolní Břežany, Czech Republic
[4] Plasma Physics Group, Imperial College London, South Kensington Campus, London SW7 2AZ, UK
[5] Department of Physics and Astronomy, University of Kentucky, Lexington, KY 40506
[6] AWE plc, Aldermaston, Reading RG7 4PR, UK

1. Introduction

It is well established that strongly photoionized plasmas are a significant aspect of accretion-powered astrophysical objects [1-6]. In such plasmas, the radiation field is intense enough to ensure that photoexcitation and photoionization rates are dominant over electron collisional excitation and ionization rates. A key variable used to characterise photoionized plasmas, the photoionization parameter $\xi = 4\pi F/n_e$ (where $F$ is the x-ray flux and $n_e$ the electron density), can reach values in excess of 1000 erg-cm s$^{-1}$ [7].

Several laboratory astrophysics experiments have attempted to recreate the relevant parameter regime, but with limited success. For example, Foord et al [8] used x-rays from the Z Machine at the Sandia National Laboratories, to both heat and decompress a solid foil. The resulting photoionized plasma achieved $\xi \simeq 25$ erg cm s$^{-1}$. In a later experiment, Fujioka et al [9] used the GEKKO-XII laser facility to achieve $\xi \simeq 6$ erg cm s$^{-1}$ with an effective radiation temperature of ~500 eV for a duration of ~ 0.2 ns. More recently, Loisel et al [10], Mancini et al [11] and Mayes et al [12] employed the Z Machine to obtain values of $\xi$ of up to $\simeq 60$ erg cm s$^{-1}$, while our experiments with the VULCAN laser in 2016 at the UK Central Laser Facility ([13] reached peak values of $\xi \simeq 45$ erg cm s$^{-1}$ with an x-ray drive duration of ~ 0.5 ns.

It is worth noting that accretion-powered x-ray sources usually involve and are modelled with a power-law spectrum, and not a quasi-blackbody spectral shape, as pointed out by other experimental teams (for example Loisel et al [10]). However, quasi-blackbody spectral distributions are common for high Z laser plasma sources. Also, a key point of what follows is that the use of such a laboratory source does allow us to control the shape of the input spectrum via, in this case, a CH filter. This means that tailored spectra can be used as input for the codes to be tested.

As noted already, the photoionization parameter is meaningful in an astrophysical context, in a steady-state situation. However, we refer to it, as a figure of merit, for comparison with previous work, but with this proviso understood. The value of this is that in designing experiments with high values of ξ we can find robust experimental tests of astrophysical codes, where photoionization plays a key role.

A key difficulty of laboratory experiments is that, in general, the electron densities are orders of magnitude greater than in astrophysical situations, thus requiring a proportionally higher x-ray flux to achieve the domination of photoionization rates over collisional rates. This leads to the need for intense pulsed x-ray sources from major facilities, with durations often in the nanosecond or shorter timescales. However, the issue then arises as to whether the plasmas created can be modelled with astrophysical codes that require the assumption of a steady-state plasma. We return to this point below.

Our previous work [13] was the first application of a novel experimental technique developed by us [14] that moves away from the use of a quasi-blackbody radiation source, in favour of one that employs keV line radiation, namely an L-shell spectrum from a Sn laser-plasma [15]. The motivation for this is that, by doing so, the role of inner-shell photoionization is enhanced relative to that due to outer-shell photoionization, and this relative importance is expected in astrophysical plasmas with a high radiation spectral temperature. Hence ensuring the importance of inner-shell ionization allows us to undertake relevant comparisons of experimental data with astrophysical code predictions, and hence permit the relevant benchmarking of such codes.

In the work presented here, we improve on our previous experiment in several ways. Firstly, by changing to the use of Ag L-shell emission, we achieve improved coupling of the x-rays to the lower ion stages of the Ar target gas, as the photon energy range is closer to the threshold of the inner-shell photoionization. This results in a significantly enhanced fluorescent signal that allows the K-β emission to be used as a diagnostic. The advantage here is that the spectral separation between emission from different ion stages is larger than for the K-α lines measured in the White et al experiment [13], and this allows emission from each ion stage to be individually resolved. In addition, the enhanced signal has allowed us to collect data at lower pressures than previously. Secondly, a problem in comparing simulations from plasma models with experimental data in our previous work was an expected significant flow of energy down the gradient of the flux, away from the laser-plasma x-ray source. This was a result of using one-sided irradiation of the target gas, with a typical inner-shell ionized electron ejected with 0.5-1 keV of kinetic energy having a range of hundreds of microns. In the experiment described in this paper, we have used double-sided irradiation which, as shown below, leads to a low gradient in flux expected at the centre of the sample. Also, the average energy of an electron ejected from the inner shell is now expected to be < 500 eV and thus the range is smaller, leading to more localised deposition of energy as assumed in modelling. Thirdly, in the current experiment, we have fielded two spherical crystal spectrometers to measure the K-α and K-β spectra with spatial resolution in two orthogonal directions.

The paper is organized as follows. In section 2 we discuss the set-up of the experiment, and the plasma models in Section 3, with the latter compared with the experimental data in Section 4. We discuss our results and draw some conclusions in Section 5.

## 2. Experimental set-up

The experiment was undertaken using the VULCAN laser at the UK Central Laser Facility during September and October 2019. Figure 1 shows the overall layout of the experiment. Octagonal gas-cell targets contained openings at each end, where Ag-coated CH foils acted as windows of thickness 3.8 μm on some shots and 18.6 μm on the remainder. The Ag layer was 467 nm thick in all cases and coated onto the outer, laser-irradiated surface of the foil. Gas cells were pre-filled with Ar gas at pressures ranging from 10-500 mbar. The Ag foils were irradiated with 3 laser beams on each side, delivering up to 460 J per side of 527 nm wavelength laser energy, in a ~1.5 ns pulse. These beams were focused to a width of approximately 300 mm on the foil target, resulting in typical peak irradiances of about $4 \times 10^{14}$ W cm$^{-2}$. Prior experiments, detailed in Singh et al [16], indicated that approximately 1% of the laser energy was emitted from the foil into the gas-fill, in the form of L-shell x-ray emission in the 3-4 keV range. Preliminary data shots where the metal layer irradiated was Al, and thus emitted no line emission at above ~ 2keV, indicated that the observed K-shell emissions, described below, were due to inner shell photoionization caused by Ag L-shell photons and not supra-thermal electrons.

In addition to the L-shell line radiation a high-Z laser plasma is also expected to generate a broad continuum due to recombination radiation, unresolved transition arrays and bremsstrahlung e.g., [17]. This often has a quasi-blackbody spectral shape, which can be roughly characterised in many cases by a spectral blackbody temperature. As we see below, this can play a key role in the experiment.

The gas cell was fitted with 25 μm thick Kapton windows through which K-α and K-β fluorescence from the photoionized Ar gas was observed. Foils were separated by 2 mm on some shots and by 3 mm on the rest, while the lip of the Al disks onto which the foils were mounted meant that the volume visible to the spectrometers was limited to a central region further than 0.5 mm from either foil. In some shots additional slits were added to limit the view to a strip 0.2 mm wide, either in the axial or radial direction.

The principal diagnostics for the experiment were a pair of spherical Bragg crystal spectrometers. One of these was fitted with a mica crystal of 150 mm radius-of-curvature, used in 4th order to operate in the spectral range of the K-α to K-β line radiation (~2950-3200 eV). This was oriented to give spatial resolution in the axial direction, which is that normal to the laser-irradiated foils. The other was a quartz (10-20) crystal, also with 150 mm radius-of-curvature, which was oriented to give spatial resolution in the radial direction, defined as that parallel to the laser-irradiated foils. The resolution of the crystals is governed by that of a spherically curved crystal [18], the detector resolution and the plasma source size. For our case, it is the latter that governs and for a source size between 2-3 mm we calculate an expected resolution in the K-β region of $\lambda/\Delta\lambda$ ~3000.

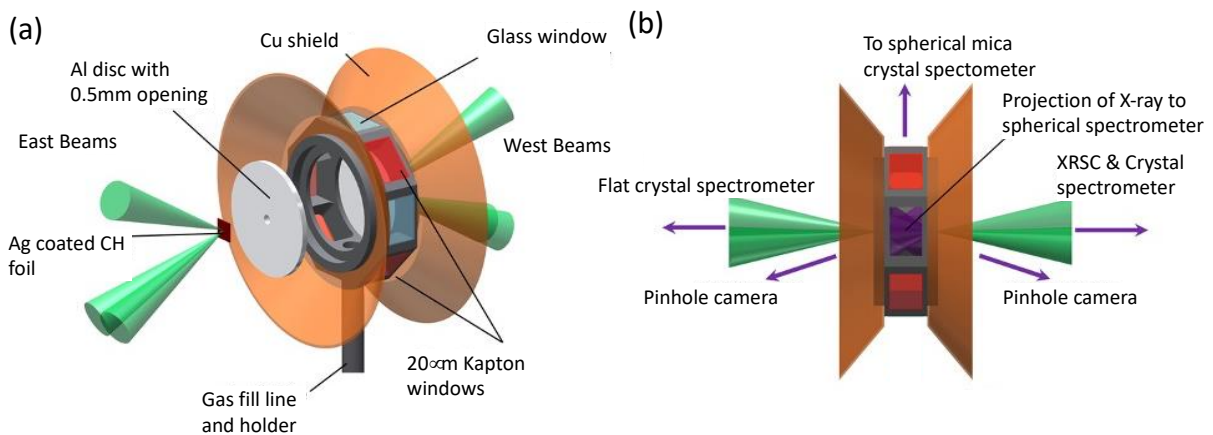

Figure 1: *(a) Partially exploded 3D view of the octagonal gas cell target and long pulse beams for our experiment on the VULCAN laser. (b) Top-down view of the target. On this view the projection of the Ag X-rays is shown. Also shown are the lines-of-sight for the diagnostics we fielded on the experiment.*

Secondary diagnostics of the experiment included two spectrometers to monitor the L-shell x-ray emission from the laser-irradiated side of each of the Ag foils. These were fitted with flat Si (111) crystals. There were also two pinhole cameras, with 10 µm pinholes, that monitored the size of the x-ray sources. In addition, the duration of the L-shell emission was monitored with an x-ray streak camera with ~100 ps resolution, viewing at an angle of around 20° from the normal to one of the target foils. Figure 2(a) shows a typical, time-integrated L-shell spectrum, and Figure 2(b) an x-ray streak camera image of the emission and a typical temporal profile, with a full-width-at-half-maximum (FWHM) of ~1.1 ns.

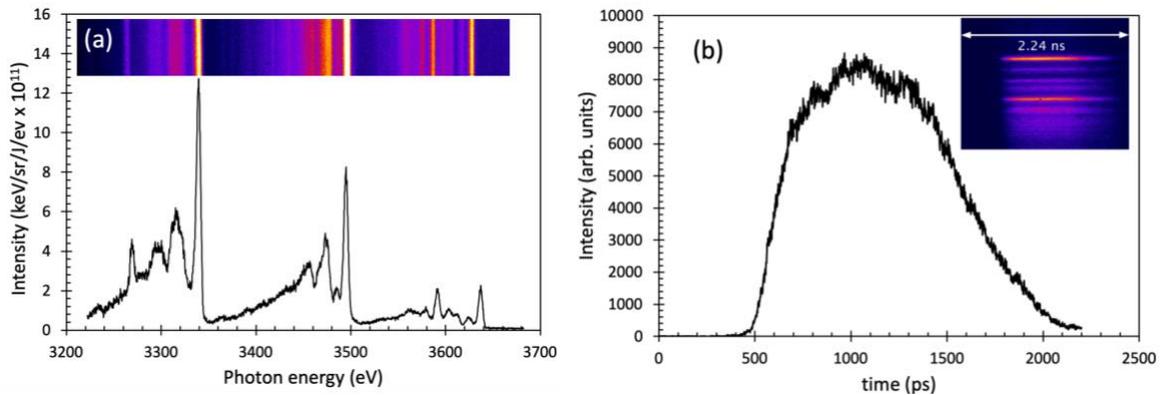

Figure 2: (a) *Typical Ag spectrum taken from the rear of a laser-irradiated foil. The spectrum is corrected for filters, CCD quantum and crystal collection efficiencies.* (b) *Typical x-ray streak camera trace, showing the duration of x-rays is of order 1 ns. The inset shows a raw, background-uncorrected data image.*

## 3. Plasma simulations

The evolution of the photoionized plasma is modelled with our time-dependent zero-dimensional computer code that assumes a local incident flux, and does not account for energy transport in or out of the local region. A geometric model based on the measured source size, conversion to L-shell and the distance between foils is used to estimate the flux within the foil. Figure 3(a) shows the on-axis peak flux expected for a 2 mm separation between foils. As in our previous work [13], collisional ionization data from Lotz [19] is fitted along with dielectronic recombination and radiative recombination rates by Shull and van Steenberg [20]. Shell-resolved photoionization cross-sections are from Verner and Yakovlev [21], and the fluorescence yields for K-shell vacancies from Kaastra and Mewe [22]. We note that using the more recent collisional ionization rates of Dere [23] leads to little difference in the history of ionization degree, due to the similarity of the cross-sections for the range of ion stages and temperatures of interest here.

After correction for filters, crystal and CCD response, the L-shell spectrum, as shown in Figure 2(a), is divided into 80, approximately equally spaced, photon groups. The expected broad band emission is modelled as a blackbody, divided into 250 photon groups up to an energy of 15 $k_B T_R$, where $T_R$ is the assumed quasi-blackbody temperature, which was typically 170 eV (~ $2 \times 10^6$ K). This quasi-blackbody spectrum was modified to account for the thickness of the CH backing of the Ag foil. The x-ray streak data were used to derive the temporal profile of the L-shell emission, and we assumed a similar profile for the quasi-blackbody emission.

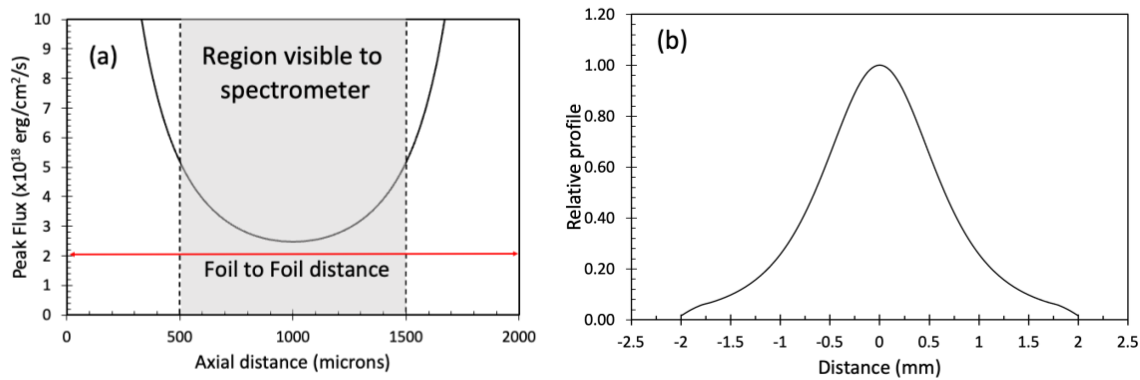

Figure 3: *Calculations with our time-dependent zero-dimensional plasma code of the expected peak incident flux of L-shell x-rays for a gas cell with 2 mm between foils. (a) On-axis case, where the shaded region is visible to the spectrometers. (b) Radial profile at the centre, between the foils, for a 300 μm flat-topped focal spot. However, we note that the profile at the gas cell centre is relatively insensitive to the focal spot profile incident on the foils.*

The electron temperature was determined by considering the energy balance between the absorbed energy from photoionization, energy lost to radiative recombination, K-α and K-β emission and bremsstrahlung (see e.g. [13]), and the internal energy contained in the plasma due to ionization potential energy and excited states. It was assumed that, compared to the timescale of the experiment, the timescales for K-shell fluorescence and Auger decay of ions with inner-shell vacancies are essentially instantaneous.

Based on the L-shell spectrum, we expect electrons ejected from the K-shell to mostly have initial energies of 100-300 eV, with a small fraction at higher values. It is assumed that electron-electron thermalisation of energy is effectively instantaneous. Standard plasma physics collision cross-sections [24] indicate that electron-electron inverse collision frequencies are up to a few 10's of picoseconds for this energy range, given an expected ion charge of Z ~ 3. Based on the simulations presented below, for our average plasma pressure conditions (50 mbar), we can expect the mean-free path for electron-ion collisions to be of order 10-100 µm for these electron energies, with mean times between collisions of up to 10 ps. This shows that energy deposition can be considered quite localised for the ~mm scale of the experiment. The electron-ion equilibration timescale $T_{eq}$ is given by,

$$\frac{dT_i}{dt} = \frac{T_e - T_i}{\tau_{eq}} \qquad (1)$$

with

$$\tau_{eq}^{-1} = 7.97 \times 10^{-11} \frac{n_e \ln \Lambda}{T_e^{3/2}} \, s^{-1} \qquad (2)$$

where $T_i$ and $T_e$ are the ion and electron temperatures, respectively, $n_e$ the electron density and $\Lambda$ the plasma parameter. As we can see from Eq. (1), the energy equilibration timescale between electrons and ions depends on the effective electron temperature. For the thermalised electron temperatures in our simulation, $T_e$ is up to ~20 eV and the timescale is around several nanoseconds, longer than the ~1 ns x-ray flux duration. Our simulations indicate, as expected, that electron-ion equilibration makes no discernible difference to our results. The ionization and recombination rates presented in Shull and van Steenberg [20] are for whole ion stages with excited states taken into account. However, excited states are also a store of internal energy that should be accounted for in determining the electron temperature from the balance of absorbed energy, sinks of energy and losses. We populate up to 20 excited states for each stage by assuming local thermodynamic equilibrium (LTE). For our electron densities of typically ~$10^{17}$-$10^{18}$ cm$^{-3}$ and electron temperatures of 10-20 eV, this is expected to be a reasonable approximation.

## 4. Comparison of experimental data and plasma simulations

### 4.1 QUB code comparisons

In Figure 4(a) we show a typical spectrum of K-β emission for one of our higher pressure (103 mbar Ar fill) gas cell experiments, obtained with the quartz crystal spectrometer, together with a lineout of the K-β spectrum, where emission from several Ar ionisation stages is clearly visible. The K-β photon energies are calculated with a Hartree-Fock atomic model, and we then use the cold K-β line energy from literature [25]) as a reference point. We note that the wavelength for the cold K-β line is calculated by the Hartree-Fock model to be within 0.01% of the experimentally observed value. The simulated spectrum shown in Figure 4(a) has been shifted slightly to match the experimental cold K-β value, and uses the experimentally measured linewidth for each ion stage. It is expected that satellite lines generated by excited states play a role in determining this. The energies of satellite lines where there is a 3s 'hole' have been calculated, and these are close to the principal transition, and hence blended due to the resolution of the spectrometer.

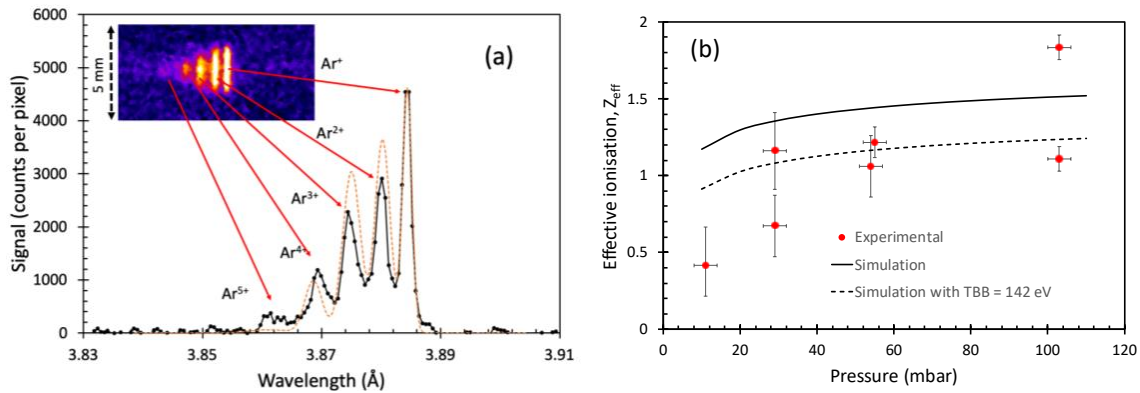

Figure 4: *(a) Experimental (black line) and simulated K-$\beta$ spectrum (dashed red line) for a shot with 103 mbar Ar fill. The experimental spectrum has spatial resolution along the direction shown by the dashed black arrow, while the $Ar^+$ line is the narrowest and indicates a spectral resolution of greater than E/DE ~2200, consistent with expectation. The radial spatial direction parallel to the target foils is indicated. (b) Comparison of effective ionization degree, $Z_{eff}$, from a series of experiments (solid points) at different pressures, with values derived from the simulated spectra generated using our plasma model (solid line). The shaded area represents a ±15% error bar in the simulation results (see text). The dashed line is a simulation assuming a reduced black-body temperature of 142 eV for the quasi-blackbody part of the input spectrum (see text).*

Since the known L-shell spectrum allows us to calculate the K-shell photoionization rates for all the ion stages, and we know the fluorescence yield for each of these [22], we can use the relative intensities of the emitted K-β lines to estimate an effective ionization state, $Z_{eff}$, and compare this to the time-integrated spectral simulation. We use the expected incident profile and forward Abel transform to calculate the line-of-sight spectrum, averaged to account for the ~ 400 μm spatial resolution of the experimental spectrum and the 1 mm wide, lateral emission region (shown in Figure 3(a)) over which the spectrometer integrates. The result is that experimentally, we obtain an average $Z_{eff}$ ~ 1.1 for the 103 mbar shot, whilst the simulation gives a higher $Z_{eff}$ ~ 1.5.

In Figure 4(b) we plot $Z_{eff}$ for different gas fill pressures, maintaining the same target and laser energy (to within 7%) as for Figure 4(a). We have derived an effective ionisation degree in the same fashion as described above. The error bars in $Z_{eff}$ for the experimental points are calculated with a Monte-Carlo approach. First, each spectral line is approximated as a Gaussian profile with three parameters: height, width, and line-centre position. This is optimised to obtain a best fit over the spectral region of interest, and the fit parameters and resultant $\chi^2$ value form the starting set of data. The fit parameters are then varied randomly within some limits, typically 30% for the peak heights and widths, and two pixels for the line centre position. If the resulting $\Delta\chi^2$ for the fit to the experimental spectrum is within the critical value for 68% confidence, we evaluate $Z_{eff}$ for that case. The critical value depends on the number of data points in the spectrum and the number of parameters fitted. We tried >$10^6$ random combinations, and evaluated the largest excursions from the best fit value of $Z_{eff}$ to determine the error bars.

In figure 4(b) we have varied the input conditions to show the sensitivity of the simulations to these, which include ±10% changes in the laser energy and assumed black-body temperature for the broad-band keV component of the drive. The latter is the strongest contributor to the variation, generating approximately a ±15% variation in the predicted ionisation. Going further, the dashed simulation curve in Figure 4(b) is calculated assuming a black-body temperature of 142 eV for the broad band contribution. This is ~15% lower than the assumed value based on limited previous experimental data and thus generates half of the expected flux. The effect on the average ionisation clearly shows that more detailed measurements in future experiments will be important to allow a better assessment of the broad band contribution. There are some key results to note from Figure 4(b). Firstly, the experimental data lie below the simulation curve at lower pressures. A second point is that higher initial pressure leads to greater ionization, seen in both simulation and experiment. This is indicative of the fact that, although the plasma heating is driven by photoionization, collisional ionization also plays a significant role in the ionization of the plasma. Simulations where the collisional and recombination rates are switched off give a simulated $Z_{eff}$ ~0.1. This defined $Z_{eff}$ is a result of integration through the plasma and the actual ionisation at the centre of the plasma is ~0.2 with collisional rates switched off.

We can see the importance of the spatial integration of the spectrometers by comparing the quartz spectrometer results, which spatially resolve in the direction parallel to the x-ray foils, to those for the mica spectrometer, which resolves perpendicular to this. For lineouts taken down the centre of each spectrum we find, experimentally, a lower effective ionisation for the mica spectrometer. This is illustrated in Figure 5, where we present the ratio of $Z_{eff}$ for the two spectrometers. We can see that our simulation predicts this trend relatively well.

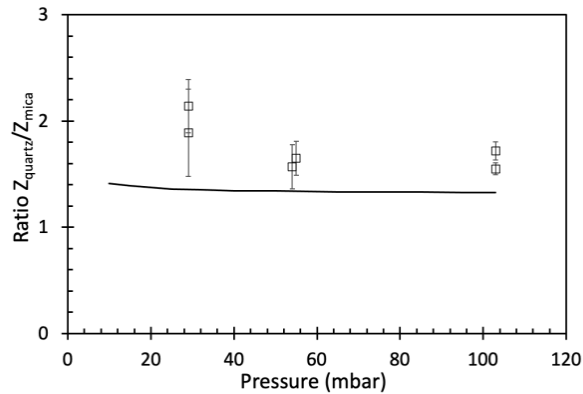

Figure 5: *Ratio of the effective average ionisation degree, $Z_{eff}$, determined from the K-β spectra for the quartz and mica spectrometers, which view the plasma with spatial integration in orthogonal directions and thus lead to different results. Data at the 11 mbar was not available for the mica spectrometer. The solid line is a prediction from our simulation.*

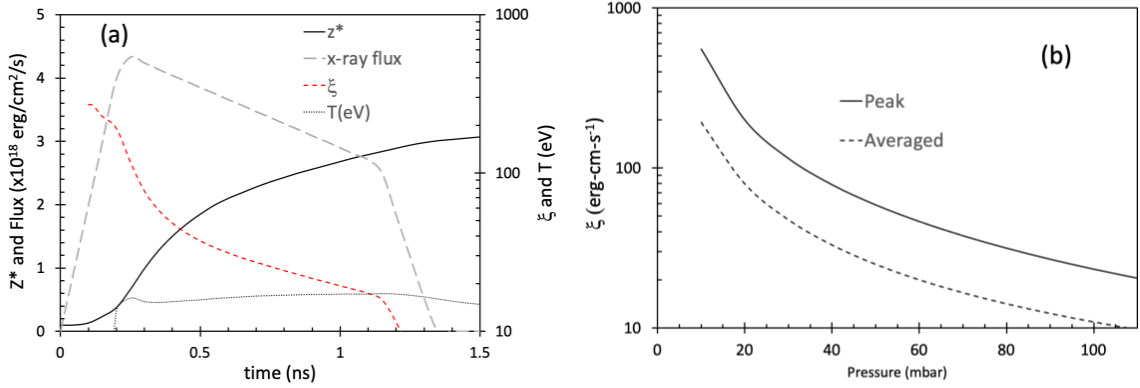

Figure 6: *(a) Simulated temporal history of mean ion charge, total x-ray flux, electron temperature and photoionization parameter, $\xi$. This simulation is for the centre of the gas cell with a 2 mm separation from the foil and a gas pressure of 29 mbar. At the peak of the x-ray flux the photoionization parameter, $\xi$ ~120 erg cms$^{-1}$. It is higher at earlier times since the electron density is much lower. As noted in the text, temporal and spatial averaging means that the emission spectrum of K-$\beta$ is characteristic of $\xi$ ~50 erg cms$^{-1}$. (b) The solid line shows the peak value of $\xi$ at the centre of the gas cell, as a function of pressure. The dashed line shows the spatially- and temporally-weighted value evaluated for the quartz spectrometer, spatially resolving in the radial direction.*

The temporal integration has a profound effect on the effective value of the photoionization parameter. In Figure 6(a) we show the temporal history of the degree of ionisation simulated at the centre of a gas cell for the 29-mbar case, the lowest pressure for which we have successful repeat shots. We can see that the total x-ray flux rises to ~4 ×10$^{18}$ erg cm$^{-2}$ s$^{-1}$ in the centre, whilst the mean ion charge increases throughout the duration of the x-ray drive, reaching Z*~3 by the end of the pulse. The photoionization parameter reaches $\xi$~120 erg-cms$^{-1}$ at the peak of the x-ray flux. However, as indicated in Figure 6(b), when we weight the photoionization parameter for K-$\beta$ emission spatially and temporally integrated for the quartz spectrometer, the spectrum is found to be characteristic of $\xi$~50 erg-cms$^{-1}$. The lowest pressure for which we obtained K-$\beta$ data (a single shot) was 11 mbar, and the emission-weighted photoionization parameter for this case would be expected to be $\xi > 100$ erg-cms$^{-1}$ based on our simulations. With a larger laser system, allowing higher signal levels, this indicates we might explore such a regime using this method.

In all shots, we obtained K-α spectra as well as for K-β. For the former, the lower ion stages are separated by only about 1 eV, which is smaller than the doublet separation and thus emission from different ion stages is blended. However, some shots were taken using a spatial slit of 0.2 mm height. By reducing the effective source size in the spectral direction, this allowed both a modest improvement in resolution to about E/DE ~2500, and the selection of emission from the most highly ionised region. This came at the cost of reducing the K-β emission, but we were able to use that from K-α in these shots to demonstrate the importance of the broad-band quasi-black-body emission in generating photoionization from the outer shells of the ions. We illustrate this effect in Figure 7 through the variation of the CH backing on the Ag foils. In a short series of shots, gas cells with a 3 mm gap between their Ag foil windows were fitted with a CH backing of either 3.8 μm or our standard 18.6 μm thickness. Since the CH layer faces the gas, the thickness of this layer affects the transmission of the softer x-rays from the Ag layer and thus the overall flux incident on the gas.

For the thicker backing the flux at the centre of the 3 mm gas cell is expected to be $\sim 2\times 10^{18}$ erg cm$^{-2}$ s$^{-1}$, with about 28% of this coming from the ~keV broadband flux, while for the thinner foils the flux increases to $\sim 5\times 10^{18}$ erg cm$^{-2}$ s$^{-1}$, with over 70% now contributed by the quasi-broadband flux. It is evident from Figure 7 that the additional softer broadband X-ray component greatly increases the degree of ionisation.

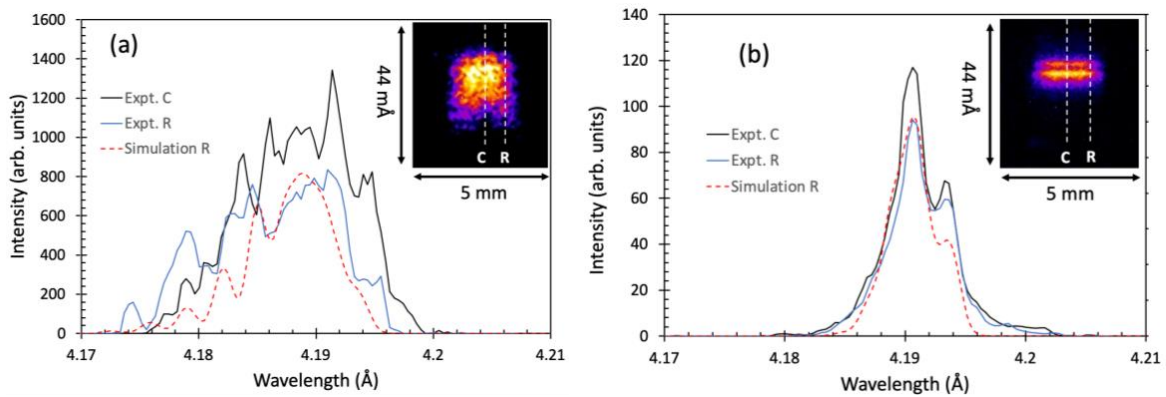

Figure 7: *K-α emission data taken with the mica spectrometer for 106 mbar gas cells with a 3 mm gap between the foils. (a) 3.8 μm CH backing and (b) 18.6 μm CH backing. The effect of an enhanced broadband X-ray flux in (a) is evident. In both cases, the lineout labelled 'R' is 0.75 mm away from the centre, while the lineout from the centre position is labelled 'C'. The simulations*

*are performed for the R position and scaled to the peak of the experimental profiles.*

This change in the broadband ~1 keV component affects two key parameters in which we are interested. Firstly, the increase in mean ion charge, due to increased broad-band flux, lowers the peak photoionization parameter from ~17 to ~12 erg-cms$^{-1}$ for the gas cells shown in Figure 7, which had a pressure of 106 mbar.  The second change is to, what we shall call, the effective spectral temperature for the photoionization source. As noted above, a key point in the discussion of Hill and Rose [14] is that the use of multi-keV photons allows an enhancement in the importance of inner-shell photoionization, which mimics the effect of a higher temperature blackbody flux. We have determined an effective spectral temperature for our shots in two different ways using our simulations. Firstly, we calculate the total photoionization rates for the K-shell and M-shells of the argon ions. For the case of 3 mm gas cells with 18.6 µm CH backing of the Ag foils, the photoionization rates for K-shell and M-shell are simulated to be roughly equal at the centre of the cell, for our expected flux of L-shell photons plus a filtered and diluted blackbody source at 170 eV spectral temperature. Also, the ratio of the K-shell and M-shell rates remains around **unity** throughout the pulse duration. Further simulations, allowing for an unfiltered black-body source, but diluted to give the same maximum flux, reproduced **this** balance between inner-shell and outer-shell photoionization as when this source was assumed to have a spectral temperature of $T_{rad}$ ~1250 eV. Thus, we define the effective spectral temperature as the black-body spectral temperature needed to reproduce the same ratio of inner-shell to outer-shell photoionization rate as for our experimental spectrum, containing L-shell photons. Corresponding calculations for the gas cells with the 3.8 µm CH backing to the Ag foils were expected to show a lower effective radiation temperature due to the stronger contribution of the broadband emission, as shown in Figure 8(b). Indeed, the ratio of M-shell to K-shell photoionization rate is now ~30 and the results indicate an effective spectral temperature of 500 eV. This large increase in the ratio is partly driven by increased flux of the total broadband component but mostly due to the increased weighting of lower energy photons closer to the outer-shell photoionization threshold. This indicates the need, in future work, to have more detailed diagnostics of the quasi-blackbody flux at sub-keV photon energies.

A second method of defining the effective spectral (colour) temperature of the drive is to model the expected temporal history of mean ion charge. As for the first method, this was performed using the experimentally determined flux and comparing to results for a diluted black body, maintaining the same flux. Our results are shown in Figure 8(a) for the case of 18.6 μm CH backing. We can see that assuming the an x-ray drive consisting solely of a black-body source with $T_{rad}$ ~700 eV but diluted by the distance from the foil to the centre of the gas cell, produces a very similar ionization history to the case where we adopt the experimentally determined L-shell and broadband emission. For the thinner 3.8 μm CH backing, an equivalent temperature of 400 eV is found through similar modelling. In the case shown in figure 8(a) the broad-band emission is a black-body at 170 eV filtered by 18.6 μm CH with a flux of $5.5 \times 10^{17}$ ergcm$^{-2}$s$^{-1}$ at the target plane. On its own, this produces a lower ionisation than that shown in figure 8(a), reaching Z* ~ 1.8 which can be reproduced by a pure black-body of ~475 eV if we are limited to the same flux.

In these comparisons we should note that fixing the flux for a black-body leads to lower ionisation for higher spectral temperatures, as the photon distribution moves further away from the outer shell photoionization thresholds. Experimentally, filtering leads to lower flux and thus ionisation, as seen in figure 7. However, the value in filtering experimentally lies in being able to tailor the spectral profile to allow a different balance between outer-shell and inner-shell photoionization.

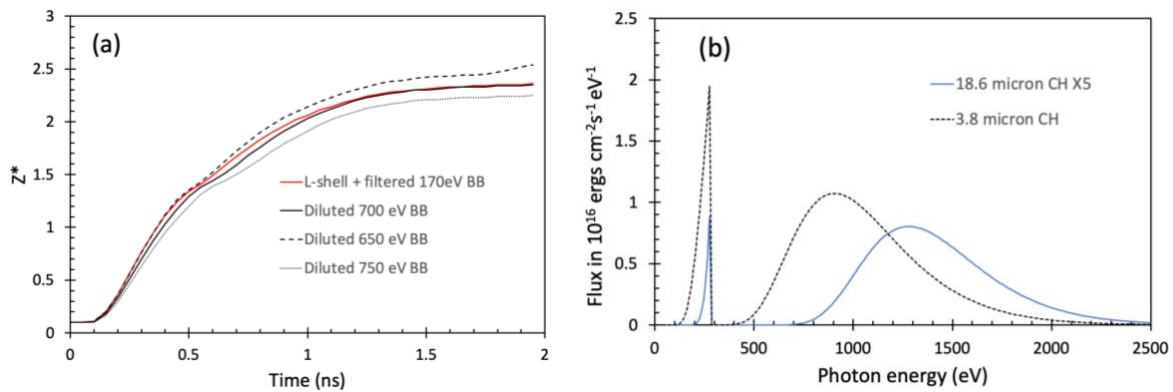

Figure 8: *(a)Simulated temporal history of mean ion charge at the centre of a 3 mm gap gas cell with 18.6 μm CH backing on source foils at 106 mbar pressure. The peak flux is $2 \times 10^{18}$ erg cm$^{-2}$ s$^{-1}$ in all cases. (b) Comparison of the expected broad-band flux at the cell centre for an assumed quasi-black body of 170 eV with the two thicknesses of CH backing.*

The in-house QUB code is zero-dimensional but is run multiple times for each position along the radial direction (direction parallel to the plane of the drive foils). This allows an Abel transform to be made to simulate the radial profile that the quartz spherical crystal can measure. In figure 9 we show a comparison of simulation against experiment for the first three K-β lines.

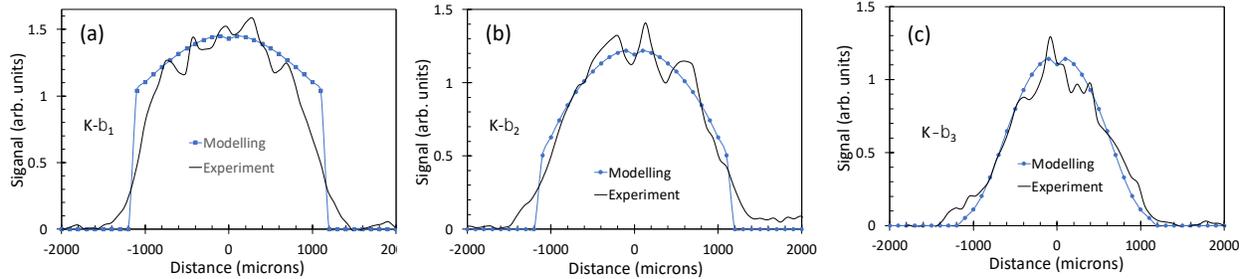

*Figure 9: Spatial profiles of the first three K-β lines compared to simulation using the QUB code. Note the cut-off due to the finite window size (2.3 mm) that prevents Abel inversion of the data. Instead a forward Abel transform of the simulation at different radial positions is used to reproduce the raw experimental profiles.*

4.2 Comparison with Cloudy code simulations

A key motivation for experiments of the type reported here is to use them as benchmarking tools for codes routinely used to model astrophysical objects. With this in mind, we have compared our simulation results to those from the Cloudy code [26-28, in particular with the time-dependent version, as this is most appropriate for experiments on nanosecond timescales. The Cloudy models were computed with a version of the code equivalent to the 23.00 release. This uses a backward Euler time integration with 0.01 ns timestep. The intensity of emission incident was contained in two tabular SEDs, one for the L-shell based on experimental measurements between 3-4 keV and in 80 photon groups, the other for the broad band as discussed for the in-house code and ranging from 10 eV to 3keV in 250 logarithmically spaced photon groups. These were varied in time in a piecewise-linear manner chosen to match the experimental illumination, and the response of the system modelled using the time-dependent capabilities of the code [28]. A parallel-beam illuminating intensity was chosen equal to the mean in the cell, rather than attempting to approximate the details of the experimental geometry. A constant 298K blackbody was also included, to ensure that the initial state of the Ar sample was cold and neutral. The Ar sample was taken to be 1 mm deep. Cloudy requires elemental abundances to be set in proportion to the H abundance, so the Ar densities were scaled relative to a (negligible) density of 1 cm$^{-3}$ of hydrogen nuclei.

In Figure 10 we show some comparisons of results from our in-house code and Cloudy models for the case of 2 mm foil separation at 103 mbar pressure and with 18.6 μm of CH between the Ag foil and the Ar gas. The simulation is for the centre of plasma, equidistant between the foils (1 mm from each) and is not spatially averaged. In Figure 10(a) we see that the electron temperature history of the two codes is broadly similar, reaching a maximum of ~1.8-2.0×10$^5$ K. For the QUB code the small drop in T$_e$ around 0.15 ns appears to be connected to the slow initial timescale for ionisation, coupled with the fact that the excited states are assumed to be instantaneously populated in LTE, thus taking up absorbed flux as internal energy. In Figure 10(b) we see that the average degree of ionisation for Cloudy is predicted to reach a slightly higher value at 2 ns, with Z* ~4 compared to Z* ~3.5 predicted from our in-house code.

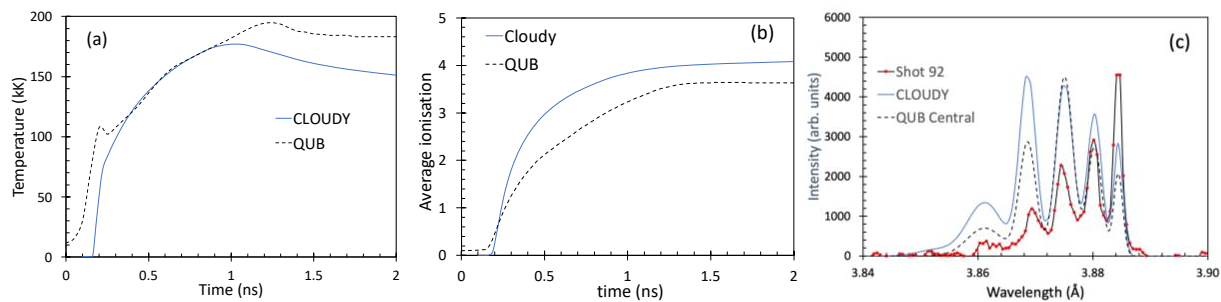

Figure 10: Comparison of time-dependent CLOUDY simulations with our in-house code for 103 mbar pressure. (a) Electron temperature in kK at the centre of the gas cell. (b) Average ionisation at cell centre. (c) Simulated K-β spectra from both codes at the cell centre compared to experimental data (Shot 92) averaged across the cell centre. Cloudy results: UK Ministry of Defence © Crown owned copyright 2023/AWE.

In figure 10(c), we show time-averaged simulated spectra that are both scaled to have the same maximum intensity. The simulated spectrum from Cloudy clearly reflects the higher mean ion charge compared to the in-house code. Although the Cloudy code is run for a single set of conditions in the centre of the gas cell, we include, for reference, the experimental data for the radially resolving spectrometer. In the next section we discuss these comparisons in more detail.

## 5. Discussion and conclusions

Our experimental data and simulations, plus their comparison, allow us to make some broad conclusions. The effective degrees of ionization derived from the experimental K-β spectra are in broad agreement with the predictions of our simulations, except at lower pressures, where the former decrease more sharply. However, the trend to higher mean ion charge at increasing pressure is seen in both experiment and simulation, indicating the importance of collisional processes. Furthermore, we note that our simulations predict that, due to the different axis of spatial resolution, the effective ionization observed by the mica spectrometer will be smaller than that by the quartz spectrometer, and this is seen experimentally.

The moderately good agreement between experiment and simulations from our plasma code gives us some degree of confidence in our assessment of the likely photoionization parameters achieved. For the lowest pressure where data were obtained (11 mbar) this reaches an emission-weighted value of >100 erg-cm s$^{-1}$, into the regime of interest for astrophysical applications. For pressures of 29 mbar, where we have repeat shots, the emission-weighted values of ~50 erg-cm s$^{-1}$ are competitive with those obtained using other experimental methods, for example the work of Fujioka et al [9], Loisel et al [10], and Mayes et al. [12]. Scaling to higher energy laser systems with > 10 kJ is feasible and, although modelling indicates that the increased flux is offset by a higher ionization and thus electron density, photoionization parameter values in excess of $\xi$ = 100 erg-cm s$^{-1}$ are entirely feasible at this pressure. Since increased L-shell flux leads to higher emission, clear, well-resolved, K-β emission at lower pressures may then allow achievable emission-weighted values higher than this.

The effect of the flux of 3-4 keV and ~ 1 keV photons on the degree of ionization has been explored experimentally and in our simulations by varying the thickness of the substrate for the Ag foils that generate the L-shell flux. An important step for future experiments will be to have a more careful determination of the x-ray spectrum and its temporal history, as currently it is inferred from a limited range of measurements in a narrow range of photon energies (1-1.2 keV) and from previous work [13]. A key conclusion, however, is that even in cases with a larger flux of softer x-rays, the effective spectral temperature (defined in Section 5 in two different ways) can be in the range of hundreds of eV. This is a major goal of this type of experiment and will allow the testing of astrophysical codes under conditions relevant to accretion-powered sources, where the plasma is dominated by a large x-ray flux. It is worth restating that these temperatures are very difficult to create in a laboratory environment, where hohlraum targets that can produce ns-duration sources are generally limited to $T_{rad}$ ~ 300 eV (e.g.,[29]) and require very large laser energies.

An important aspect of our work has been to compare our own in-house simulations with those from the time-dependent Cloudy code. The latter is run for a single set of conditions and hence we compare the predicted conditions at the centre of the gas cell rather than spectra. However, it is clear from the comparison that our experiments are suited to generating data that can be usefully compared to astrophysical codes, where the flux and spectral content are in a relevant parameter space. There is broad agreement between simulations in terms of temperature reached and degree of ionisation. However, there is clearly much scope for future work to explore, for example, the effect of rates used and how excited states are included and dealt with.

A comparison of simulations from the QUB and Cloudy codes with argon K-experimental data reveals quite good agreement in the case of the QUB predictions, but Cloudy overestimates the intensities of the higher ionisation species. However, as noted above, Cloudy is run for a single set of plasma conditions, while our experimental data are averaged across the cell centre. Hence the discrepancy with Cloudy predictions is not surprising. In future experiments, we plan to obtain both temporally and spatially resolved experimental spectra at high resolution, which will allow a more reliable comparison with Cloudy.

A key issue to be addressed in the future is the desirability of creating an experiment that can reasonably be compared with existing codes that assume steady-state conditions. We can see from Figures 5 and 7 that the timescale for the evolution of average degree of ionization is of nanosecond order. In fact, if we define a characteristic timescale by $\tau^{-1} = z^{-1} \partial z / \partial t$, where z is the mean ion charge, then this varies from ~ 0.2 ns at the peak of the x-ray flux to ~3 ns at the end. An obvious possibility to improve on this is to go to longer pulse laser systems. For the conditions predicted in Figure 5, we expect a typical sound speed of ~$10^4$ ms$^{-1}$ and thus, for a mm scale system, a decompression time of ~100 ns. This justifies the omission of hydrodynamic motion in our simulations and demonstrates the scope for future work using longer pulse x-ray drives. However, use of a longer x-ray pulse at the same time as maintaining the desired photoionization parameter is likely to require the use of a more powerful laser facility than the VULCAN one employed in the experiments discussed here.

In summary, laboratory astrophysics experiments are usually limited to reproducing one or two key characteristics of the physics of astronomical sources. In our case these are the photoionization parameter and effective spectral temperature relevant to accretion-powered objects. However, we believe there is scope in experiments of the kind described here to contribute to a wider investigation of the processes occurring in accretion-powered sources, including the testing of codes used to model these systems. Indeed, the research will be relevant to any plasma where photoionization is significant, whether they are astrophysical in nature or related to Earth-bound applications such as laser-fusion capsules.

## 6. Data Access Statement

The reduced and analysed data published in this paper are available from the corresponding author at d.riley@qub.ac.uk on request. These, plus the raw data on which the paper is based, have been stored on the Queen's University Research Data Management System, and once again are available from d.riley@qub.ac.uk on request.

## 7. Acknowledgments

This work was supported by the UK Science and Technology Facilities Council through grant ST/P000312/1. We would like to thank the UK Central Laser Facility staff who run the laser, target area and target preparation facilities for their contributions.